\documentclass[twocolumn,superscriptaddress,showpacs,amsfonts,amsmath,amssymb,natbib, jap]{revtex4-1}
\usepackage{graphicx}
\usepackage{dcolumn}
\usepackage{bm}
\usepackage{subfig, textcomp}

\begin{document}

\title{A road to hydrogenating graphene by a reactive ion etching plasma}

\author{M. Wojtaszek}
	\email{M.Wojtaszek@rug.nl} 
	\affiliation{Physics of Nanodevices, Zernike Institute for Advanced Materials, University of Groningen, Groningen, The Netherlands}
\author{N. Tombros}
	\affiliation{Physics of Nanodevices, Zernike Institute for Advanced Materials, University of Groningen, Groningen, The Netherlands}
\author{A. Caretta}
		\affiliation{Optical Condensed Matter Physics, Zernike Institute for Advanced Materials, University of Groningen, Groningen, The Netherlands}
\author{P. H. M. van Loosdrecht}
		\affiliation{Optical Condensed Matter Physics, Zernike Institute for Advanced Materials, University of Groningen, Groningen, The Netherlands}
\author{B. J. van Wees}
	\affiliation{Physics of Nanodevices, Zernike Institute for Advanced Materials, University of Groningen, Groningen, The Netherlands}

\date{\today}

% keywords: plasma hydrogenation, graphene, Raman spectroscopy, graphane

\begin{abstract}
We report the hydrogenation of single and bilayer graphene by an argon-hydrogen plasma produced in a reactive ion etching (RIE) system. Electronic transport measurements in combination with Raman spectroscopy are used to link the electric % electric or electronic?%
 mean free path to the optically extracted defect concentration. We % also?% 
 emphasize the role of the self-bias of the graphene in suppressing the erosion of the flakes during plasma processing. We show that under the chosen plasma conditions the process does not introduce considerable damage to the graphene sheet and that hydrogenation occurs primarily due to the hydrogen ions from the plasma and not due to fragmentation of water adsorbates on the graphene surface by highly accelerated plasma electrons. 
For this reason the hydrogenation level can be precisely controlled. The hydrogenation process presented here can be easily implemented in any RIE plasma system.     
\end{abstract}

\pacs{Valid PACS appear here} \maketitle
%%%%%%%%%%%%%%%%%%%%%%%%%%%%%%%%%%%%%%%%%%%%%%%%%%%%%%%%%%%%%%%%%%%%%%%%%%%%%%%%%%%%%%%%%%%%%%%%%%%%%%%%%%%%%%%%%%%%%%%%%%%%%%%%%%%%%%%%%%%%%%%%%%%%%%%%%%%%%
\section{\label{sec:Introduction}Introduction}

Hydrogenation of carbon materials, e.g. graphite, carbon nanotubes or carbon foams, has triggered a large technological and scientific interest, with its main focus on hydrogen physisorbtion in hydrogen storage systems \cite{Strobel2006}. However, for electronic applications the chemisorption of hydrogen is even more interesting, as it allows for tuning of electronic properties in carbon conjugated systems. An excellent candidate for such manipulation is graphene, a single layer of graphite, built only from sp$^{2}$ carbons and demonstrating high carrier mobilities \cite{Dean2010}. Similarly to single wall nanotubes \cite{Kim2004}, the small volume and large contact area of graphene makes %chemisorbtion => %
chemisorption of hydrogen an efficient way to modify its electronic properties \cite{Boukhvalov2008, Gao2011}. Depending on the H coverage one can tune the transport properties of graphene from metallic to semiconducting, and ultimately to an insulating state for its fully hydrogenated derivative {\em graphane} \cite{Elias2009}. Opening of a bandgap by hydrogenation in otherwise gapless graphene can be also an elegant way to fabricate a circuit consisting of a single material: graphene, with both metallic and semiconducting parts. 

Apart from microelectronic applications, the influence of hydrogen on electronic transport in graphene has great scientific relevance as well. In particular to understand the role of localized defects as scattering centers limiting carrier mobility \cite{Ni2010}, the transition in charge transport from the Drude type (in pristine graphene) to the variable range hopping type (in strongly hydrogenated graphene) \cite{Elias2009}, or in predictions of magnetism originating from hydrogen defects \cite{Yazyev2007, Zhou2009}. 

Chemisorption of hydrogen on a graphene surface changes  the carbon electronic orbitals from $sp^{2}$ to $sp^{3}$ hybridization and results in a localized state. The potential barrier for hydrogen adsorption to the surface of graphene is about 0.2 eV \cite{Boukhvalov2008, Ruffieux2002}. Part of this energy is consumed by the displacement of the carbon out from the graphene plane to obtain the tetragonal $sp^{3}$ geometry. This adsorbtion barrier is lower in initially curved or protruding structures (at grain boundaries, lattice defects or on ripples), where structural deformation is already present  \cite{Ruffieux2002}. 
For effective and controllable hydrogenation of graphene several techniques have been explored so far, including exposure to an atomic hydrogen source \cite{Balog2009, Katoch2010, Sessi2009},\  electron beam (e-beam) exposure of highly hydrated lithography resist HSQ \cite{Ryu2008} and e-beam exposure of a water adhesive layer on graphene \cite{Yavari2010, Jones2010a}. Among this techniques exposure to an argon-hydrogen plasma produced in a DC \cite{Elias2009} or RF source \cite{Luo2009} seems promising alternative due to the high energy and reactivity of the incident hydrogen ions, what enables their chemisorption even on the flat surface of graphene. Hydrogenation by an Ar/H$_{2}$ plasma can lead to a high and fast hydrogen uptake, it does not require special sample preparation and is compatible with microfabrication techniques. 

Estimation of the H content in micromechanically cleaved graphene flakes after hydrogen treatment is very difficult. The standard methods known from graphite, like thermal programmed desorption (TPD)  \cite{Denisov2001, Nechaev2004, Zecho2002}, are insensitive to the possible amounts of desorbed hydrogen from micro-sized flake. Estimation of H coverage from scanning tunneling microscopy (STM) topography images carries limitations that STM probes the surface only locally, measurements are time consuming and difficult when graphene is deposited on the insulating substrate. An appealing alternative is  Raman spectroscopy, which is a relatively easy, non-destructive, non-contacting and quick method to probe H coverage from even micrometer sized samples and can be carried out at room temperature and atmospheric pressure. Chemisorption of H induces Raman bands which are normally symmetry forbidden in the graphene spectrum. The assignment of these bands to hydrogen adsorbates allows an indirect estimate of the H content \cite{Malard2009}. 

In this work we demonstrate the hydrogenation of graphene by an RF plasma of an argon-hydrogen gas mixture using reactive ion etching (RIE). This technique has not been explored for graphene hydrogenation so far, despite the fact that RIE is widely used for electronic device microfabrication. 
We characterize the hydrogenation properties of the RF plasma and its reversibility under moderate thermal annealing by means of Raman spectroscopy. Further we present the electronic transport measurements in single layer (SLG) and bilayer graphene (BLG) which enables us to relate the structural defects to graphene transport properties.  
 In the control experiment we compare the effect of the Ar/H$_{2}$ plasma with the pure Ar plasma in two types of sample (in bare flakes on insulating substrate and in graphene devices). The observed differences highlight the role of the floating potential of the non-contacted graphene flakes for acceleration of the graphene erosion. As this effect is completely suppressed in graphene devices, we conclude that there graphene hydrogenation happens primarily due to the hydrogen ions and not to highly accelerated plasma electrons fragmenting water add-layer on graphene surface, as suggested in Ref. \cite{Jones2010} .           

\section{\label{sec:Results} Experimental methods}
\subsection{RF plasma conditions}
\begin{figure}
\centering
\includegraphics[width=0.95\columnwidth]{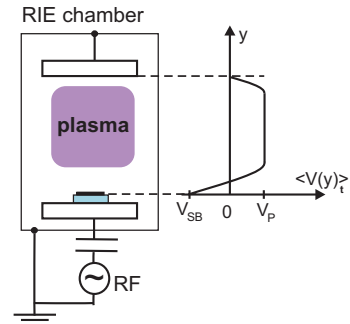} 
\captionsetup{format=hang,justification=centerlast, font=small}
\caption[]{(Color online) Schematic drawing of the reactive ion etching chamber used in the experiment. The sample is placed on the bottom electrode, connected to the RF power source, while the top electrode and the walls are grounded.  On the right the corresponding potential profile during the plasma operation. The positive bias V$_{P}$ inside the plasma and negative self-bias V$_{SB}$ at the sample are indicated. Graphene can be insulated from the bottom electrode by the SiO$_{2}$ substrate or connected to it by fabricated metallic contacts. In latter case graphene will have the same self-bias as the bottom electrode. }
\label{fig:RIE_chamber}
\end{figure}

The  hydrogenation is performed  in a reactive ion etching reactor with a parallel plate geometry, schematically depicted in Fig. \ref{fig:RIE_chamber}. The diameter of the bottom electrodes, on which the samples are placed, is 300~mm  while the opposite top wall of the chamber serves as a grounded counter-electrode. A high frequency generator operating at 13.56~MHz is capacitively coupled to the bottom electrode, and a matching of the electrical network to the plasma is accomplished by mechanical tuning of the impedance in the circuit.  

In view of safety considerations, we use a gas mixture of H$_{2}$ (15\%) with Ar (85\%) as a balance gas. The ionization energy of Ar, E$_{Ar}$~=~15.76~eV, is very close to the ionization energy of H$_{2}$, E$_{H_{2}}$~=~15.42~eV, therefore the induced plasma is composed of ions from both species.  The inlet of gas is controlled by an Ar mass flow controller. In all presented plasma hydrogenation processes the gas flow is kept constant at 200~sccm and the pressure in the chamber is 0.05~mbar. To reduce the reactivity of the plasma, especially carbon sputtering by Ar ions, we use the plasma at the lowest ignition power, P~=~3~W (power density is $\sim4$~mW/cm$^{2}$), and we tune the circuit impedance to reduce the built-in DC self-bias between the bottom electrode and the plasma (V$_{SB}$ in Fig.~\ref{fig:RIE_chamber}) down to zero. 
 Reducing V$_{SB}$ to zero minimizes the ion acceleration and possible sputtering effects on graphene. We analyze two cases, one where the graphene flakes are electrically insulated from the chamber electrodes (by the SiO$_{2}$ substrate) and one where the flake is in electric contact with the source electrode (latter called a graphene device). In the first case, the potential of the flake is floating, which may result in negative charging of the flake before the plasma quasi-equilibrium state and V$_{SB}$=0 is reached (in the first 3 seconds after the plasma ignition). This charging is largely suppressed for graphene device, which is in electrical contact with the chamber electrode.     
 On the basis of work of Nunomura {\em~et~al.} \cite{Nunomura2007} we estimate that we are in a collisional regime, with ion bombardment energy in the range of 5-20~eV and that the dominant hydrogen radicals are $H_{3}^{+}$, with a much smaller concentrations of $H_{2}^{+}$ and $H^{+}$. We note that the processing conditions and hydrogenation speed are different from the one explored in Ref. \cite{Luo2009}. The gas pressure in that process is 2 orders of magnitude higher than it is here, and Luo {\em et al.} used a grounded bottom electrode.

\subsection{Raman spectroscopy of prisitine and hydrogenated graphene}
Information about the H content can be obtained indirectly from Raman spectra \cite{Luo2009}. In pristine graphene only two vibrational modes are Raman active: an in-plane optical vibration of  $E_{2g}$ symmetry, at 1580~cm$^{-1}$, called G band, produced by $sp^{2}$ carbon network and a resonantly enhanced two phonon scattering process, around 2670~cm$^{-1}$, called 2D (or sometimes G'). The presence of $sp^{3}$ defects breaks the translational symmetry in the graphene lattice  and activates other resonant transitions. The most significant is so-called defect band D at 1340~cm$^{-1}$, forbidden in the ideal $sp^{2}$ graphene lattice. The D band results from a second order process involving intervalley elastic scattering of electron by defect and inelastic scattering on phonon. It is worth noting that the 2D mode is an overtone of the D peak, with the difference that in case of the 2D band an electron is scattered by a second phonon instead of a defect. Additionally $sp^{3}$ defects induce a much weaker D' band at 1620~cm$^{-1}$, coming from intravalley defect scattering and a peak, which can be assigned to the combination of the D and G mode (G+D) at $\sim2940$~cm$^{-1}$ \cite{Malard2009}. These properties of graphene make Raman spectroscopy a sensitive tool for detection of chemisorbed H defects. It is worth noting that the physisorbed molecules do not change the hybridization of carbons and hence do not contribute to the Raman signal of the D band. 
In the Ar/H$_{2}$ plasma process presented here, one has to take into account also the effect of the Ar ions, which by bombarding graphene could induce other $sp^{3}$-type defects: vacancies. These defects also contribute to the D band intensity in Raman spectra; therefore, care must be taken when one assigns the D band intensity solely to the H adatoms.  Later in this work we prove by studies of thermal desorption that the sputtering effect of Ar ions is largely suppressed in the chosen plasma conditions (considerably low RF power, high gas pressure). This assures the hydrogen origin of $sp^{3}$ defects. To quantify the level of hydrogenation we use the integrated intensity ratio $I_{D}/I_{G}$ of Raman bands, which relates the amount of $sp^{3}$ defects in the graphene lattice to its inherent $sp^{2}$ bonds. 
Raman spectra are obtained using a Horiba T64000 micro-Raman spectrometer with 532~nm laser excitation wavelength, spectral resolution of $\sim2$~cm$^{-1}$, laser spot size $<$10~$\mu m$ in diameter and power density below  0.5~mW to avoid laser induced heating. First, we study the evolution of the D band and its amplitude in comparison with the G band in Raman spectra at various plasma exposure times. For that purpose we select a set of graphene flakes deposited on SiO$_{2}$/Si substrate (300~nm of SiO$_{2}$) by micromechanical cleavage of Kish graphite. For each flake we obtain a pristine Raman spectrum. With that we exclude the presence of initial disorder. By analyzing the shape and FWHM of the 2D band  we confirm the number of layers in the chosen flakes \cite{Graf2007, Ferrari2006}. Then each sample is exposed separately to the Ar/H$_{2}$ plasma for a specific amount of time and immediately after that the Raman spectrum in ambient conditions is acquired. 
\captionsetup[subfloat]{captionskip=-2.2em,margin = 0.1em,justification=raggedright,singlelinecheck=false,font=normalsize, position=top}

%places caption for subfigure like (a), (b), (c), without extra text, on top, with small overlay on figure
\begin{figure}[ht!]
\centering
  \subfloat[]{  
   \includegraphics[width=0.7\columnwidth]{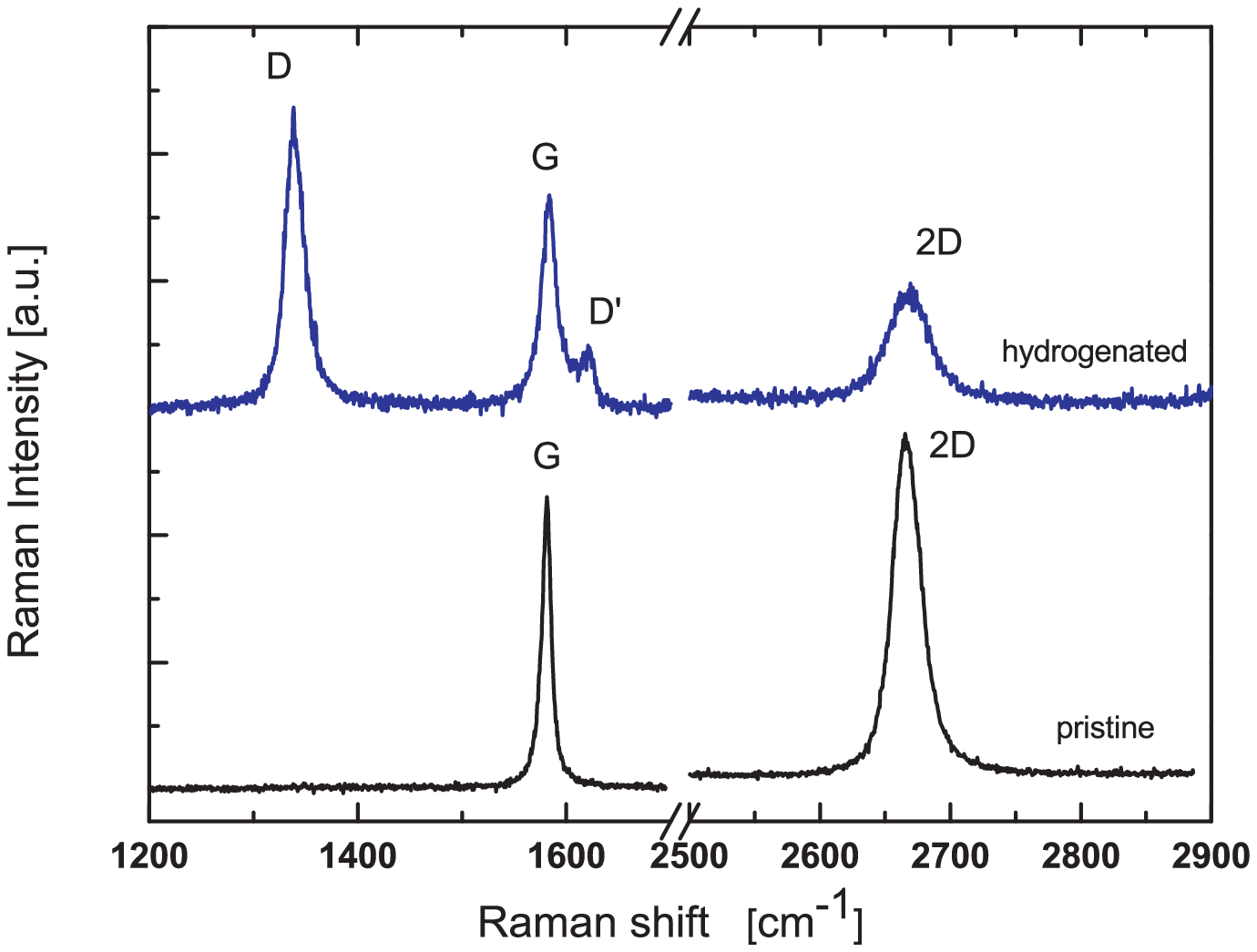} 
   \label{fig:Raman_example}}
     \qquad 
  \subfloat[]{  
   \includegraphics[width=0.7\columnwidth]{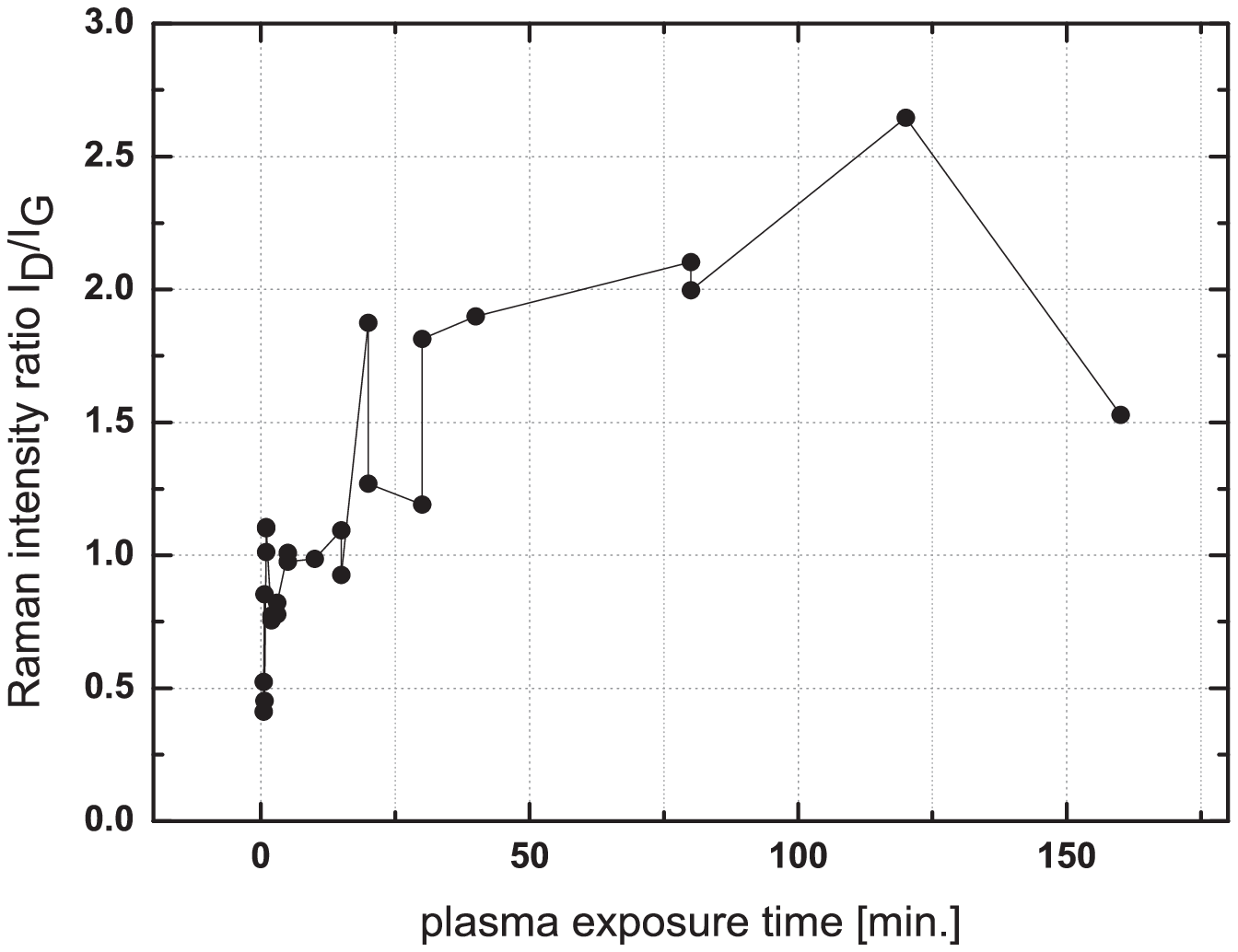} 
   \label{fig:Raman_afterH}}
     \qquad 
  \subfloat[]{  
   \includegraphics[width=0.7\columnwidth]{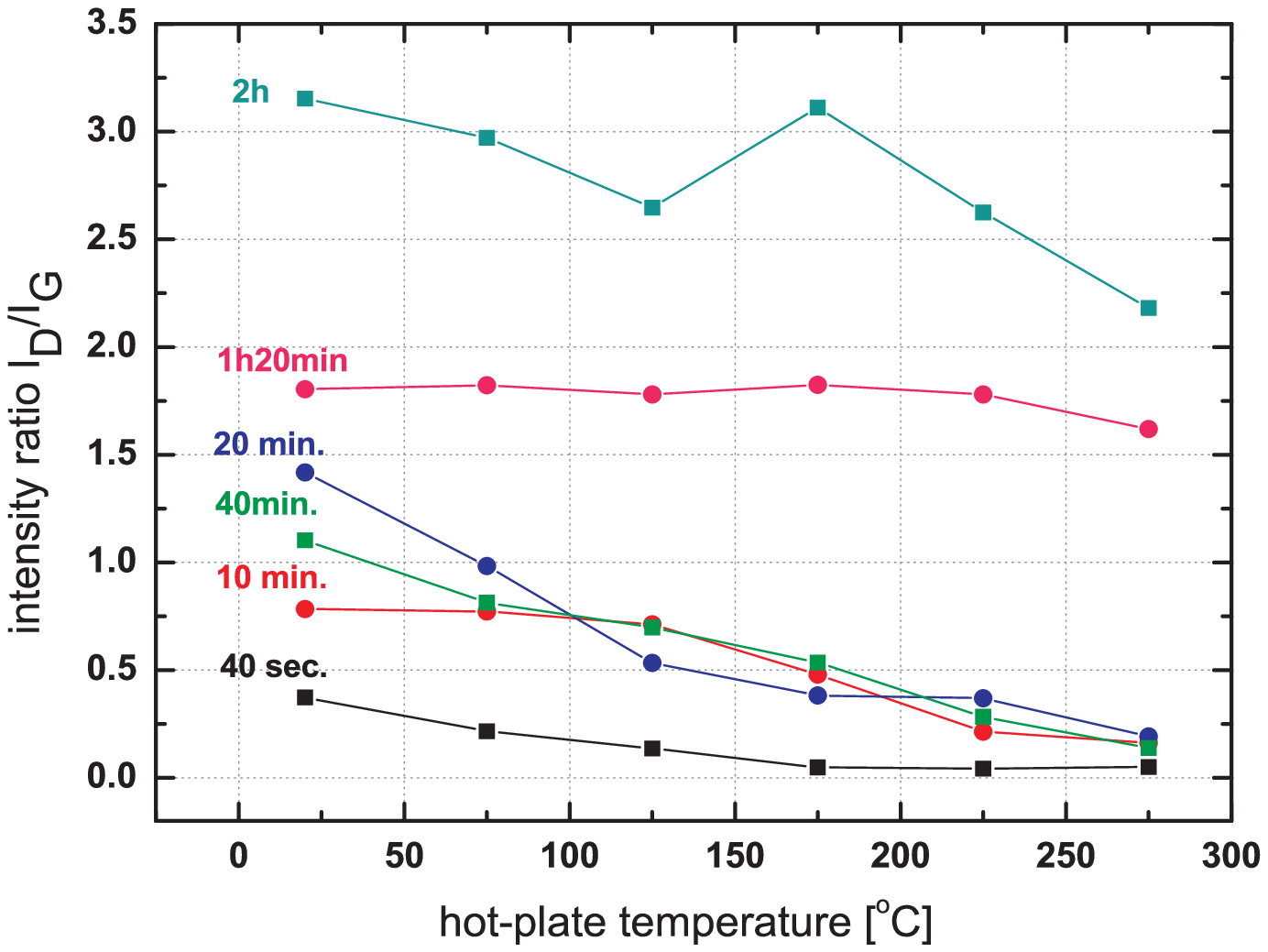} 
   \label{fig:Raman_heating}}

\captionsetup{format=hang,justification=centerlast, font=small}
   
\caption[]{(Color online) (a) Raman spectrum of pristine single layer graphene (black) and after 20~min of exposure to the Ar/H$_{2}$ plasma (blue). Exposure induces additional Raman bands: a D band around 1340~cm$^{-1}$ and a weaker D' band around 1620~cm$^{-1}$. The increase of FWHM of original graphene bands (G, 2D) is apparent. (b) Integrated intensity ratio between the D and G bands of SLG after different Ar/H$_{2}$ plasma exposure times. The scattering of the data for different samples is attributed to the floating potential of the graphene flake during exposure. (c) The change of the $I_{D}/I_{G}$ ratio of exposed flakes under annealing on hot-plate for 1~min. The plasma exposure time for each flake is indicated next to the corresponding $I_{D}/I_{G}$ values. In flakes exposed for less than 1~h the D~band could be almost fully suppressed ($I_{D}/I_{G}<$0.2), which confirms the H-type origin of defects. In longer-exposed samples (80 min and 2 h) annealing does not significantly reduce $I_{D}/I_{G}$, which suggests a different nature of defects there, e.g. vacancies. }
\end{figure}

A typical Raman spectrum of graphene before and after the plasma exposure is shown in Fig.~\ref{fig:Raman_example}. Hydrogenation results in activation of additional vibrational modes, two of which are depicted in Fig.~\ref{fig:Raman_example}: a disorder induced D band at $\sim1340$~cm$^{-1}$ and D' band at 1620~cm$^{-1}$. The evolution of the integrated intensity ratio between the D and G bands in the Raman spectrum, $I_{D}/I_{G}$, after different exposure times is shown in Fig.~\ref{fig:Raman_afterH}. We note its similar behavior to that presented in Ref. \cite{Luo2009}. The increase of the exposure time results in the increase of the ratio between the D and G band up to the point where there are so many defects in the graphene lattice that the graphene electronic band structure is degraded, reducing possible optical transitions for both D and G bands \cite{Ferrari2007}.  The initial increase and then decrease of the $I_{D}/I_{G}$ ratio with an increasing number of defects in graphene is reported irrespective of the origin of the defects \cite{Lucchese2010,Moser2010, Lin2010, Childres2011}. After hydrogenation, all original Raman bands of graphene show an increase of their FWHM, which is attributed to the local deformation of the lattice and a larger variation in vibrational/phonon energy.

\subsection{Reversibility of hydrogenation under annealing}

To confirm that the defects in graphene detected by Raman spectroscopy originate from H adsorbates, we study the change of the $I_{D}/I_{G}$ ratio after heat treatment. The comparative studies of hydrogen desorption in graphite by TPD show that H starts to desorb already at moderate temperatures, $>$100~\textcelsius, with the desorption maxima at 175~\textcelsius \ and 290~\textcelsius \  and estimated activation energy for desorption is 0.6~eV \cite{Zecho2002}. Note that these temperatures are too low to heal possible vacancies in graphene. We perform the heating in a nitrogen environment on a hot-plate, with temperature ranging from 75~\textcelsius \  to 275~\textcelsius, each time for 1~min. As can be seen in Fig.~\ref{fig:Raman_heating}, heating results in a decrease of the ratio $I_{D}/I_{G}$. It starts already at 75~\textcelsius \  and continues decreasing with increase of heating temperature. Desorption of hydrogen below 100~\textcelsius, also reported in \cite{Luo2009}, can originate from different nature of hydrogenation by plasma in comparison with an atomic hydrogen source, as energetic ions can bind to graphene in more diverse, also meta-stable, configurations of hydrogen clusters.

After heating at 275 \nolinebreak \textcelsius, $I_{D}/I_{G}$ drops below 0.2 in the case of the samples exposed to plasma for less than 1~h. The samples exposed for 80~min and 2 h show a much smaller decrease of defect band intensity with temperature. This means that after prolonged exposures the D band in these flakes must originate primarily from carbon vacancies rather than H adsorbates. In a control experiment we expose the graphene flakes to the pure Ar plasma at the same RIE exposure conditions. We observe that the pure Ar plasma induces substantial etching of graphene, with complete erosion of the flake after about 30~min. The different etching rate of the Ar/H$_{2}$ plasma versus pure Ar plasma can be explained by the mass difference between H and Ar ions. Lighter H ions are faster accelerated by the bias difference between the plasma and graphene and they reach the graphene surface sooner than Ar ions. By charge transfer H ions effectively neutralize the negative potential of the flake, reducing the self-bias voltage between the sample and the plasma and the acceleration of much heavier Ar ions. 
Although the carbon vacancies seem to contribute substantially to the D band signal after the plasma exposures with Ar, this effect is completely suppressed in graphene devices, where the flake is in electric contact with bottom electrode. Exposure of the contacted flake to the Ar plasma did not produce any defect related Raman bands even after prolonged exposure ($>$3~h) and later in the text we show no significant change in graphene electronic mobility under the Ar plasma exposure. 
This emphasizes the role of the floating potential of the graphene sample for amplifying the etching speed.

\subsection{Electronic transport in hydrogenated graphene}
\begin{figure}[ht!]
  \subfloat[]{    
   \includegraphics[width=0.7\columnwidth]{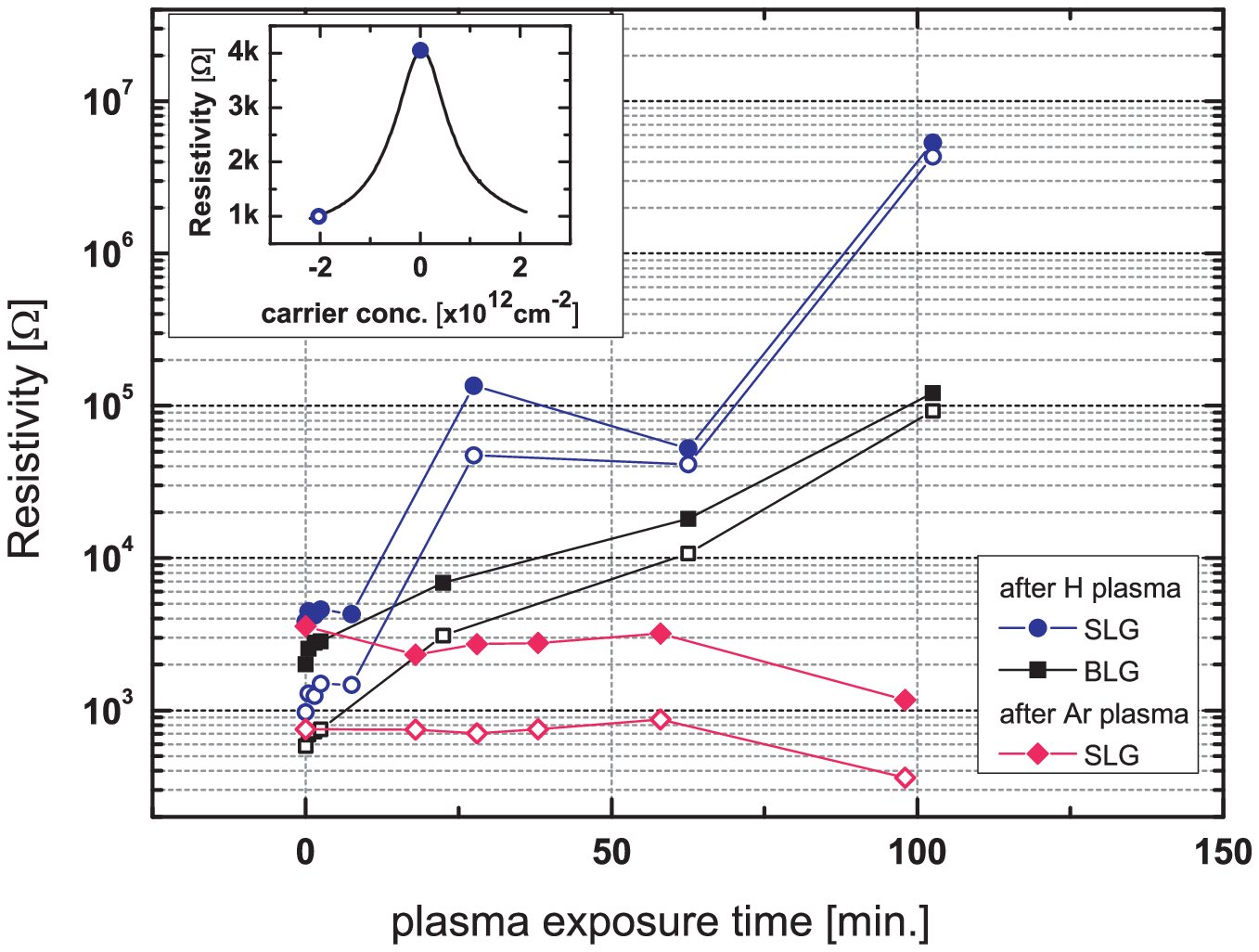} 
   \label{fig:Rho_vs_plasma_time}}
   \quad
  \subfloat[]{  
   \includegraphics[width=0.7\columnwidth]{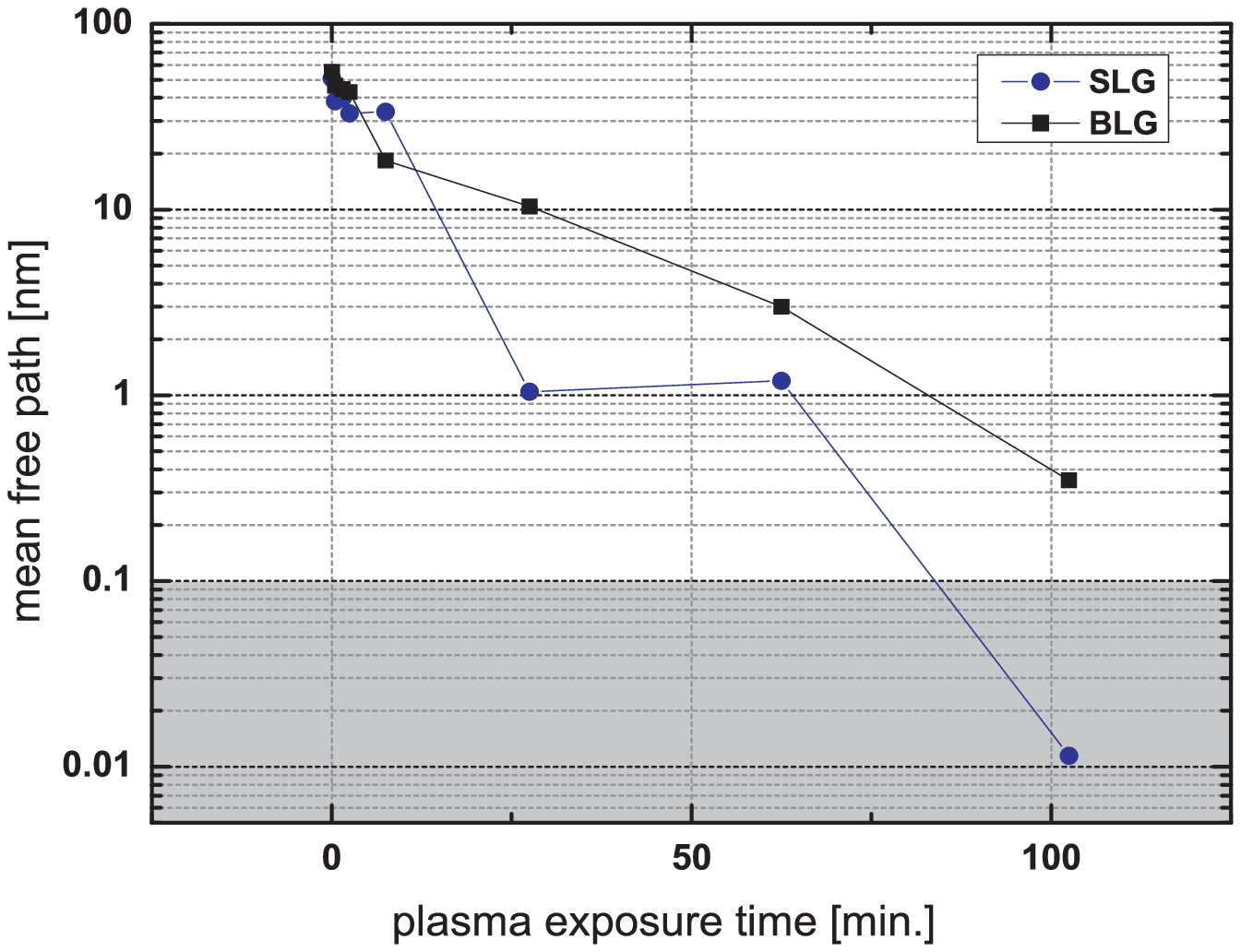} 
   \label{fig:mfp_vs_plasma_time} }
\captionsetup{format=hang,justification=centerlast, font=small}
\caption{(Color online) (a) Resistivity of single (blue dots) and double layer graphene (black squares) after several exposures to the Ar/H$_{2}$ plasma. Filled circles represent the resistivity at the Dirac point, open circles represent the resistivity in a metallic regime (at $2 \times 10^{12}$ cm$^{-2}$ carrier density). For comparison, filled and open diamonds describe the resistivity changes in SGL after the Ar plasma exposure. The inset presents the exemplary resistivity curve for SLG. (b) Mean free path of charge carriers in graphene after the exposures. The shaded area indicates the values below the length of C-C bond, where the calculations of the mean free path loses its physical meaning.  }
\end{figure}

To gain more information about the role of different H coverage on electronic transport, we perform 4 terminal resistivity measurements in single and bilayer graphene devices after sequential exposure to the Ar/H$_{2}$ plasma (devices are exposed simultaneously). The measurements are done at room temperature and in vacuum shortly after the plasma exposure. 
The inset of Fig.~\ref{fig:Rho_vs_plasma_time} shows exemplary resistivity measurements at different charge carrier concentrations for SLG device. The carrier concentration $n$ can be extracted from the charge induced by the gate voltage V$g$ with respect to the voltage of the charge neutrality point (CNP) V$_{D}$ (also called Dirac point, where the valence band of graphene touches the conduction band) by using the formula: $n = C_{g}/e (V_{D}-V_{g})$, where gate capacitance C$_{g}$=115~aF/$\mu m^{2}$ for 300~nm SiO$_{2}$.
Upon exposure the position of the Dirac point shifts to positive voltages, indicating the hole doping from H. Linking this shift directly to the amount of adsorbed H is however not appropriate here, as the measurements are done {\em ex-situ} and other dopants, like physisorbed water molecules, could screen the doping induced by H \cite{Leenaerts2008}. 
For that reason we focus on the resistivity changes at the charge neutrality point and in a high doping regime, where graphene shows metallic behavior (here arbitrarily taken at $\sim2 \times 10^{12}$~cm$^{-2}$). In Fig.~\ref{fig:Rho_vs_plasma_time} one can see that with the increase of the exposure time the SLG resistivity changes from a few k$\Omega$ to M$\Omega$ \  and for BLG to hundreds of k$\Omega$. Upon hydrogenation the resistivity difference between CNP and a high doping regime changes from $\sim$3~k$\Omega$ to $\sim$300~k$\Omega$, and its gate voltage characteristic broadens indicating the large amount of charge impurities/inhomogeneities. (If one defines the width of resistivity dependence $\rho$ from the charge carrier concentration as the distance between its deflection points, then upon hydrogenation this width changes in SLG from  $8\times 10^{11}$~cm$^{-2}$ to  $>1\times 10^{14}$~cm$^{-2}$). As one might expect, the increase of graphene resistivity with exposure time is slower for BLG than for SLG, as there the graphene layer underneath is unexposed. Moreover, BLG shows a monotonic increase of resistivity with exposure, whereas for SLG we observe a non-monotonic change in resistivity, which suggests a change in the transport mechanism for exposure times $>$30~min. The same behavior is reflected by the electron mean free path $l$, calculated here using the formula: $l=2D/v_{F}$, where $v_{F}$ is Fermi velocity of electrons in graphene, $v_{F}=10^{6}$~m/s, and $D$ is a diffusion coefficient (obtained from Einstein relation $D=\sigma/e^{2} \nu$, $\nu$ is the density of states). In the calculation we neglect the effect of finite temperature on the density of states (DOS) and any broadening due to charge impurities; the interlayer coupling in DOS of bilayer graphene $\gamma_{1} = 0.4$~eV, after \cite{Ohta2006}). Figure \ref{fig:mfp_vs_plasma_time} shows a change of the mean free path $l$ %which % 
with the H plasma exposure. It decreases monotonically for BLG and non-monotonically for SLG. The shaded area marks the mean free path distances below the length of the C-C bond ($\sim$1.4~\AA{}), where the diffusion transport model loses its physical meaning. The fact that the estimated mean free path for SLG after $\sim$2 h of exposure enters this range provides us with evidence that the transport there can no longer be described by the semi-classical Drude model. Low temperature measurements presented in Ref.~\cite{Elias2009} show that in the heavily hydrogenated samples the transport enters a variable range hopping regime, but the full description of this transition is still lacking. 

 Additionally, in a control experiment we perform the same electrical characterization of graphene devices exposed to the pure Ar plasma. The change of graphene resistivity upon exposure is confronted with the effect of Ar/H$_{2}$ treatment in Fig.~\ref{fig:Rho_vs_plasma_time}. We see that after Ar exposure graphene resistivity does not change, remaining in the k$\Omega$ regime and also no D band could be resolved in Raman spectra. With these two characterization techniques we measure no influence of the Ar plasma on the graphene devices in spite of strong graphene erosion in the case of non-contacted flakes (such flakes are completely sputtered after 30~min). This confirms that with the chosen plasma conditions no detectable damage is introduced by Ar ions and that in the flakes with zero self-bias the defects detected by Raman spectroscopy come only from H adsorbates. These findings also disprove the suggestion of Ref.~\cite{Jones2010} that under exposure to the Ar/H$_{2}$ plasma the observed defect band in Raman comes from the fragmentation of a water add-layer by high energy plasma electrons. If that were the case, we should see the Raman band after exposure to Ar in graphene devices even when the Ar plasma does not introduce defects itself. The high energetic plasma electrons from Ar ions should similarly fragmentate a water add-layer, which is always present in the vicinity of graphene due to the used substrate (SiO$_{2}$ is hydrophilic). Since no Raman band is observed after Ar exposure, the Ar plasma does not cause graphene erosion and that water layers do not contribute to hydrogenation in the plasma process described here.

\subsection{Relation between the mean free path and defect density}

\begin{figure}[ht!]
\includegraphics[width=0.7\columnwidth]{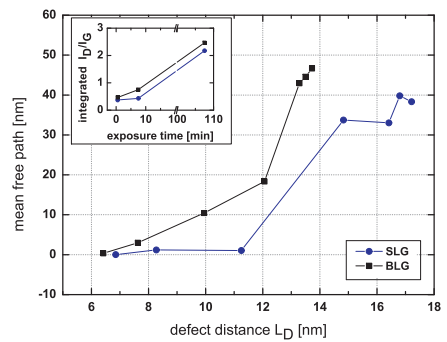}
\captionsetup{format=hang,justification=centerlast, font=small}
\caption{(Color online) Nonlinear correspondence between estimated defect distance and the calculated mean free path in single and bilayer graphene. The inset presents the measured $I_{D}/I_{G}$ ratio for SLG and BLG devices at three different exposure times. }
\label{fig:mfm_vs_defect_distance} 
\end{figure}

Having ascertained that the defects characterized by Raman spectra originate only from H, we can relate the mean free path to the defect distances L$_{D}$ extracted from the $I_{D}/I_{G}$ ratio. The commonly used Tuinstra-Kroenig experimental dependence \cite{Tuinstra1970}, which relates the $I_{D}/I_{G}$ ratio to the size of graphite nanocrystalinities and therefore defect distances, was obtained from X-ray diffraction measurements. Estimation of defect concentration from that relation is inappropriate here, because in Tuinstra-Koenig experiment only the edge defects and not the whole surface area contribute to the Raman scattering. We therefore apply a relation established for low energy (90 eV) argon ion bombarded graphene from Ref.~\cite{Lucchese2010}, which in the regime measured electronically here ($I_{D}/I_{G}<$2.5) has a form $I_{D}/I_{G}=(102\pm 2)/L^{2}_{D}$. The proportionality coefficient was  obtained experimentally for Raman laser wavelength $\lambda$ = 514.5~nm, which is close to the one used here ($\lambda$ = 532~nm) and therefore we neglect its possible energy dispersion \cite{Luo2011}. As we measure Raman spectra after 3 different exposures (see inset in Fig.~\ref{fig:mfm_vs_defect_distance}), the $I_{D}/I_{G}$ ratios for the exposures in between are estimated assuming their linear increase in time between the consecutive ratios. The estimated defect distance is compared to the electronic mean free path extracted from transport measurements in Fig.~\ref{fig:mfm_vs_defect_distance}. We observe a nonlinear relation between the defect distances and the  mean free path in both SLG and BLG.  
Assuming a parabolic dependence of the mean free path from defect distance: $l=L_{D}^{2}/\sigma$  we obtain a scattering cross section $\sigma$ of 7~nm for SGL and 4~nm for BLG. This confirms that the cross-section for electron scattering on the impurity potential is larger than the size of the structural disorder caused by this impurity. These scattering cross-section is roughly the same within the first four exposures and then it strongly increases, suggesting a coalescence of the hydrogenated regions. The lower scattering cross-section in BLG supports the theoretical predictions that the impurity potential is screened more effectively in BLG than in SLG \cite{DasSarma2010}. 
After the last exposure, the H coverage determined from the defect distance $L_{D}$ is $\sim$0.05\%.

As in Ref. \cite{Luo2009} we find that after the Ar/H$_{2}$ plasma exposure the $I_{D}/I_{G}$ ratio for BLG device is larger than that for SLG device (see inset in Fig.~\ref{fig:mfm_vs_defect_distance}). This observation is in contradiction to the Raman ratios after exposure of graphene to atomic H and when other defects are introduced \cite{Ryu2008, Ni2008}. It is also counterintuitive, as in the bilayer the presence of the second graphene layer reduces the rippling imposed by the amorphous SiO$_{2}$ substrate, which should increase the potential barrier for chemisorption of H. Also the intensity of the G band in the case of BLG should be greater than in SLG, as bilayer resting on a substrate can absorb H only on the top layer, leaving the layer beneath intact. With the same surface disorder, the $I_{D}/I_{G}$ ratio for BLG is estimated to be 3.5 times smaller than for SLG \cite{Ryu2008}. From that we conclude that the
binding of H in our process is effectively 4 times larger for BLG than for SLG. The observed discrepancy may be inherent to the reactivity of $H_{3}^{+}$ ions, the most dominant hydrogen-based component in RF plasma, and to their dissociation mechanisms at the graphene surface. Details of this process, together with the exact evolution of the $I_{D}/I_{G}$ ratio with the number of exposed layers, need computational verification.
Monte Carlo simulations of graphite bombarded with H atoms predict that the highest adsorption rate is for H-beam with incident energy of 5~eV; then in a higher energy range (around 15 eV) the H atoms are reflected back from the surface and at even higher energies ($>$30~eV) H atoms are able to penetrate through the hexagonal ring and initiate chemical sputtering \cite{Ito2008}. Here the plasma ion kinetic energy ranges from 5~eV to 20~eV, which covers both: chemisorption and reflection regime for H ions. This may explain the somewhat longer exposure times for similar hydrogenation levels than in Ref. \cite{Luo2009}. This also indicates that the efficiency of this process may be still further improved, by for example increasing the gas pressure or by increasing the RF power. Although the maximum hydrogenation limit is not explored here, this plasma technique is expected to allow a much higher hydrogen uptake than the one reported here (0.05\%).

\section{\label{sec:Conclusions}Conclusions}
In this work we report for the first time the realization of graphene hydrogenation in reactive ion etching (RIE) system.  We study the evolution of the intensity ratio of Raman bands $I_{D}/I_{G}$ and on this basis quantify the induced disorder. With moderate heating we are able to reverse the hydrogenation to almost initial level, which confirms that the observed disorder in Raman spectra stem from adsorbed H. We emphasize here the importance of graphene electric potential during the plasma exposure to suppress erosion of the flakes. We perform electrical studies of single and bilayer graphene after several plasma exposures and link them with the amount of the structural disorder characterized by Raman spectroscopy. The nonlinear correspondence between the mean free path and the estimated defect distances is highlighted, from which the scattering cross-section for hydrogen defect is obtained. We prove that under the chosen plasma conditions, hydrogenation occurs primarily due to the hydrogen ions and not due to fragmentation of a water add-layer by highly accelerated plasma electrons. We also demonstrate that by controlling the electric potential of the graphene during the plasma exposure we suppress the sputtering of carbons in graphene. For that reason the hydrogenation level can be precisely controlled and reversed. The described hydrogenation process can be easily implemented in any RIE system, which we believe will stimulate the research of hydrogenated and functionalized graphene.

%\section{\label{sec:Acknowledgement}}
\begin{acknowledgments}
We would like to thank B.~Hesp, B.~Wolfs, J.~Holstein and S.~Bakker for technical support and T.~Maassen for critically reading the manuscript. We acknowledge financial support from the Ubbo Emmius program of the Groningen Graduate School of Science, the Zernike Institute for Advanced Materials and the  Netherlands Organization for Scientific Research (NWO-CW) through a VENI grant.
\end{acknowledgments}
%
%

% 
%
%\bibliographystyle{}
%\bibliography{references}
% Note the lack of whitespace between the commas and the next bib file.

% %\end{thebibliography}

\end{document}